\documentclass[a4paper,11pt]{article}
\usepackage{pos}
\usepackage{graphicx}
\usepackage{mathtools}
\usepackage{amsmath,amsfonts,amsthm}
\usepackage{array}
\usepackage{multirow}
\usepackage{url}

\title{Sensitivity of CTA to gamma-ray emission from the Perseus galaxy cluster}
 \ShortTitle{CTA - DM in Perseus}

\author*[a, b]{Judit P\'erez-Romero}

\affiliation[a]{Instituto de F\'{i}sica Te\'{o}rica UAM-CSIC, Universidad Aut\'{o}noma de Madrid, C/ Nicol\'{a}s Cabrera, 13-15, 28049 Madrid, Spain}
\affiliation[b]{Departamento de F\'{i}sica Te\'{o}rica, M-15, Universidad Aut\'{o}noma de Madrid, E-28049 Madrid, Spain}

\forColl{CTA Consortium}  

\emailAdd{judit.perez@uam.es}

\abstract{In these proceedings we summarize the current status of the study of the sensitivity of the Cherenkov Telescope Array (CTA) to detect diffuse gamma-ray emission from the Perseus galaxy cluster. Gamma-ray emission is expected in galaxy clusters both from interactions of cosmic rays (CR) with the intra-cluster medium, or as a product of annihilation or decay of dark matter (DM) particles in case they are weakly interactive massive particles (WIMPs). The observation of Perseus has been proposed as one of the CTA Key Science Projects. In this contribution, we focus on the DM-induced component of the flux. Our DM modelling includes the substructures we expect in the main halo which will boost the annihilation signal significantly. We adopt an ON/OFF observation strategy and simulate the expected gamma-ray signals. Finally we compute the expected CTA sensitivity using a likelihood maximization analysis including the most recent CTA instrument response functions. In absence of signal, we show that CTA will allow us to provide stringent and competitive constraints on TeV DM, especially for the case of DM decay.}

\FullConference{37$^{\rm{th}}$ International Cosmic Ray Conference (ICRC 2021)\\
		July 12th -- 23rd, 2021\\
		Online -- Berlin, Germany}

\begin{document}
\maketitle

\section{Introduction}

Galaxy clusters are the largest gravitationally-bound objects in the Universe, with masses between $M_{200}\approx 10^{14}-10^{15} \rm M_{\odot}$. Dark matter (DM) is expected to account for about 80\% of their mass~\cite{RevModPhys.77.207}, while the rest is baryonic matter in the form of galaxies, gas and dust in the intra-cluster medium (ICM). Although clusters are supposedly virialized objects, the presence of hot gas, strong magnetic fields, galaxies and Active Galactic Nuclei (AGNs) produces turbulence phenomena and complex baryonic feedback in the ICM. All these astrophysical processes act as acceleration mechanisms, leading to production of cosmic rays (CR). The presence of CRs has been confirmed through observations of diffuse synchrotron emission produced by the CRs electrons and positrons at different wavelengths~\cite{vanWeeren:2019vxy}. However, the gamma-ray emission, expected either from neutral pion decay from accelerated CRs~\cite{Pinzke_2010} and/or from DM annihilation or decay of Weakly Interacting Massive Particles (WIMPs) particles~\cite{Cirelli:2010xx}, has so far avoided detection~\cite{Ackermann:2013iaq}.

Galaxy clusters are considered excellent and complementary targets for gamma-ray DM searches because of the large amount of DM they are expected to host. Given the large number of known clusters, it is key to determine which of them meet the most appropriate conditions to be searched for in gamma-rays:
\begin{itemize}
    \item Proximity to Earth ($z<0.1$), so that substantial DM-induced fluxes are expected.
    \item High-mass clusters with well-determined masses, since the annihilation flux is proportional to the mass.
    \item Low gamma-ray flux from conventional astrophysical processes~\cite{Jeltema:2008vu}.
\end{itemize}
In~\cite{S_nchez_Conde_2011} authors studied how the most promising galaxy clusters compared to the known dwarf Spheroidal galaxies (dSphs) in terms of their annihilation fluxes. Following results of DM-only N-body simulations, the substructures in the main DM halo where for the first time included for the computation of the annihilation flux. This substructures, usually called subhalos, are a natural prediction within the standard $\Lambda$CDM structure formation scenario. The authors concluded that galaxy clusters are competitive targets in terms of the annihilation flux, as long as the DM annihilation happening in these subhalos is properly taken into account.

The Cherenkov Telescope Array (CTA) will be the next generation of Imaging Atmospheric Cherenkov Telescopes (IACTs). CTA's energy sensitivity will improve up to one order of magnitude the sensitivity of current generation IACTs in the 20 GeV to 300 TeV energy range~\cite{CTAConsortium:2018tzg}, will have 2.5 better field of view and 5 times better angular resolution. CTA will have two different arrays: the Northern array in La Palma (Spain), and the Southern array in the Atacama desert in Chile. Having an excellent spectral and angular resolution, CTA has superb capabilities to perform gamma-ray DM searches, especially those focused on WIMPs.

Among the local clusters that fulfill the requirements mentioned before, the Perseus cluster is the brightest in the X-ray sky. It hosts two central AGNs, NGC1275 and IC310. Being one of the most massive clusters in the nearby Universe, Perseus is one of the most promising clusters to be detected in gamma-rays~\cite{Aleksi__2010}. More precisely, given its position in the sky, it meets the optimal conditions for observation by the CTA Northern array~\cite{CTAConsortium:2018tzg}. All these attributes make the Perseus science case an excellent discovery opportunity for CTA~\cite{CTAConsortium:2018tzg}.

\section{Dark Matter modelling in Perseus}

The expected annihilation flux in gamma-rays from the Perseus cluster can be computed as follows~\cite{Bertone:2004pz}:

\begin{equation}\label{eq:flux}
 \frac{d\Phi_{\gamma}}{dE}(\Delta\Omega, l.o.s, E)=\frac{d\phi_{\gamma}}{dE}(E)\times\;J(\Delta\Omega, l.o.s),
\end{equation}

where $\frac{d\Phi_{\gamma}}{dE}$ is the gamma-ray annihilation/decay flux, $\frac{d\phi_{\gamma}}{dE}$ is the particle physics term containing the spectral information \cite{Cirelli:2010xx} and $J$ is the so-called astrophysical factor. The particle physics term can be written, assuming Majorana WIMPs, in terms of the selected annihilation/decay channel ($dN^{PP}/dE$), the mass ($m_{DM}$) and the averaged annihilation cross section ($< \sigma v>$) or the DM particle lifetime ($\tau_{DM}$) as

\begin{equation}\label{eqn:annihil-flux}
 \frac{d\phi^{Annihil}_{\gamma}}{dE} = \frac{<\sigma v>}{8\pi m_{DM}^2}\frac{dN^{PP}}{dE},
\end{equation}
\begin{equation}\label{eqn:decay-flux}
\frac{d\phi^{Decay}_{\gamma}}{dE} = \frac{1}{4\pi m_{DM}\tau_{DM}}\frac{dN^{PP}}{dE}.
\end{equation}

The astrophysical factor or J-factor (D- for decaying DM) accounts for the DM distribution in the halo, parametrized in the DM density profile $\rho_{DM}$:

\begin{equation}
 J(\Delta\Omega, l.o.s) = \int_{\Delta\Omega}\int_{l.o.s}\rho_{DM}^2,
\end{equation}
\begin{equation}
 D(\Delta\Omega, l.o.s) = \int_{\Delta\Omega}\int_{l.o.s}\rho_{DM}.
\end{equation}

From the X-ray surface brightness data of the Perseus galaxy cluster, one can derive the cluster's main parameters. We build the DM density profile for the main halo starting from its measured mass. In our study, we adopt the mass estimate in~\cite{Aharonian:2017rqy}. We also need to assume a DM profile, in this case, the Navarro-Frenk-White (NFW) profile~\cite{Navarro:1995iw, Navarro:1996gj}:

\begin{equation}\label{eqn:NFW}
 \rho_\mathrm{NFW}(r)=\frac{\rho_0}{\left(\frac{r}{r_s}\right)\left(1+\frac{r}{r_s}\right)^{2}}.
\end{equation}

The two parameters in the NFW profile are computed by making use of the concentration-mass ($c-M$) relation in~\cite{Sanchez-Conde:2013yxa}. The obtained DM density profile parameters for Perseus are given in Table~\ref{tab:clusters_mass_models}.

\begin{table}[h!]
\begin{center}
\scalebox{1}{
\begin{tabular}{|c|c|c|c|c|c|c|}
\hline
Cluster & ($l$, $b$) & $D$  & $M_{200}$  & $R_{200}$ & $\rho_{\rm 0}$ & $r_{\rm s}$  \\
& [deg] & [Mpc] & $[10^{14}\; \mathrm{M}_{\odot}]$ & [$10^2\;$kpc]     & $[10^6\;\rm M_{\odot}\,kpc^{-3}]$  & [$10^2\;$kpc]    \\
\hline
Perseus & (150.57, -13.26) & 75.01   & $7.52$     &   $18.65$     &       $1.20$       & $3.71$  \\ 
\hline
\end{tabular}}
\caption{\label{tab:clusters_mass_models} DM density profile parameters for the Perseus galaxy cluster assuming the NFW profile and the $c-M$ relation in \cite{Sanchez-Conde:2013yxa}.}
\end{center}
\end{table}

\begin{figure}[h!]
\centering
\includegraphics[width=0.95\linewidth]{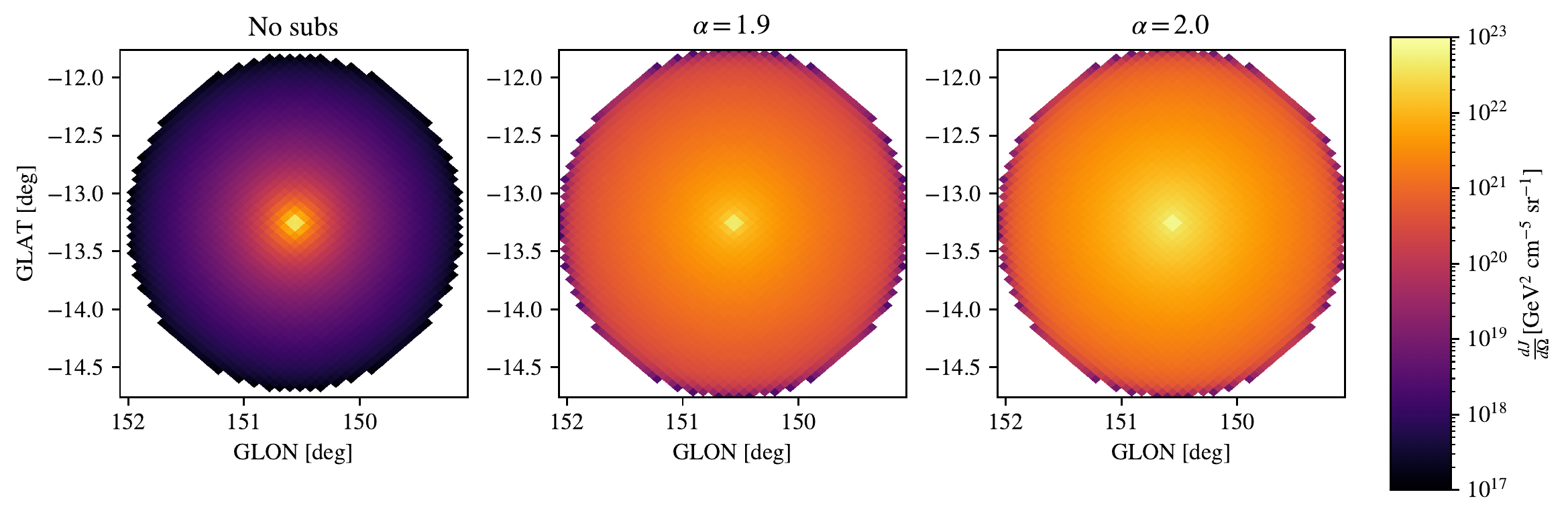} 
\caption{Two-dimensional spatial templates showing the expected spatial morphology of the DM annhilation flux in Perseus. The inclusion of different amounts of substructure as described in the text alters the flux especially in the outskirts of the cluster.}
\label{fig:templates}
\end{figure}

We also include in the modelling the expected substructures that contribute to the DM-induced gamma-ray flux. We note that this is only relevant for the annihilation case. We assume an NFW density profile also to describe the internal structure of subhalos. We parametrize the population of subhalos inside the main halo via three basic factors: radial distribution (SRD), mass distribution (SHMF) and subhalo concentration. Following the state-of-the-art of the subhalo DM population as given by N-body simulations, we will adopt:
\begin{itemize}
    \item SRD: Anti-biased distribution following the results of the Via Lactea - II Milky-Way-sized cosmological simulation~\cite{Diemand:2008in}.
    \item SHMF: $\frac{dN}{dM}\propto M^{-\alpha}$, where $\alpha$ can take two values $\alpha = 1.9, 2.0$; following~\cite{Springel:2008cc} and~\cite{Diemand:2008in}, respectively. 
    \item Concentration: We use the parametrization specifically derived for subhalos developed in~\cite{Moline:2016pbm}.
\end{itemize}
In our work, we define three different models of the subhalo population in Perseus. With the definition of these three models, we will be able to provide an estimate of the uncertainty on the exact configuration of halo substructure in Perseus. Indeed, we will obtain conservative values (no substructure model) for the DM annihilation flux and also upper bounds ($\alpha = 2.0$ model). To compute the J- and D- factors we use the \texttt{CLUMPY} free software \cite{Charbonnier:2012gf, Bonnivard:2015pia, Hutten:2018aix}. The results of the DM modelling for annihilation and decay are shown in Table \ref{tab:Js} and the corresponding differential annihilation flux profiles are represented in Figure \ref{fig:templates}. 

\begin{table}[h!]
\centering
\begin{tabular}{| c | c || c | c |}
\hline
Annihilation &  $\log_{10} J$ [GeV$^2$cm$^{-5}$] & Decay &  $\log_{10} D$ [GeV cm$^{-2}$]\\
\hline
$ J_{T}$ & 17.42 & \multirow{3}{*}{$D_{T}$} &  \multirow{3}{*}{19.20} \\
\cline{1-2}
$ J_{T}^{1.9}$ & 18.48 & & \\
\cline{1-2}
$ J_{T}^{2.0}$ & 18.93 & & \\
\hline
\end{tabular}
\caption{\label{tab:Js} Annihilation of the different DM models (no substructure, SHMF $\alpha=1.9$, SHMF $\alpha=2.0$) and decay results for Perseus, integrated up to $R_{200}$ (symbolized as the sub-index $T$). See text for for details.}
\end{table}

From the J-factor values of Table \ref{tab:Js} we can notice the effect of taking into account the substructure in the cluster. This enhancement of the J-factor when including the subhalos in the DM modelling is known as the \textit{subhalo boost factor} $B$, defined as $B = 0$ for the case where no subestructure is included. For the model with $\alpha=1.9$ the \textit{boost factor} is $B=10.5$, and for $\alpha=2.0$ we obtain $B=31.4$.

\section{CTA sensitivity to DM-induced gamma-ray from Perseus}

\begin{figure}[h!]
\centering
\includegraphics[width=\linewidth]{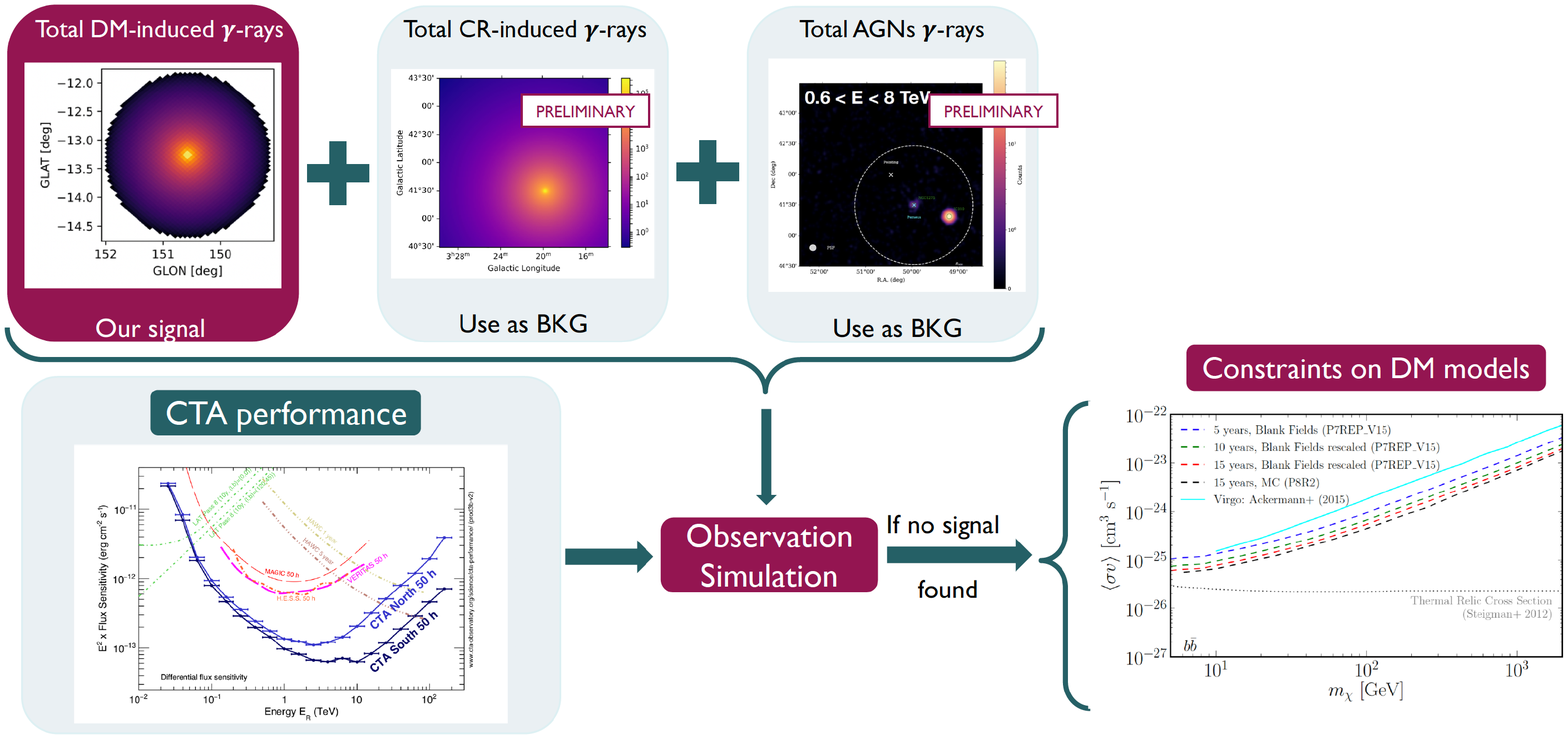}
\caption{Sketch of the analysis pipeline to compute the sensitivity of CTA to gamma-ray induced DM signal. The bottom right plot is only shown as an example of constraints from this type of searches, and was extracted from \cite{Charles:2016pgz}.}
\label{fig:roadmap}
\end{figure}

We use the most up-to-date Instrument Response Functions (IRFs - \texttt{prod3b-v2}) to compute the prospects for observing gamma-rays from Perseus using CTA. Besides, we also use the two available softwares to perform the analysis, \texttt{gammapy}\footnote{\url{https://docs.gammapy.org/0.18.2/index.html}} and \texttt{ctools}\footnote{\url{http://cta.irap.omp.eu/ctools/}}. In our analysis, we include DM annihilation/decay as gamma-ray source. In future steps of the project, we will consider also CR-induced gamma-ray emission and the emission from the two AGNs hosted in Perseus. An illustrative sketch of the analysis pipeline to follow in our work is shown in Figure \ref{fig:roadmap}.

The analysis strategy is based on the widely used ON/OFF observation method~\cite{Li1983AnalysisMF}. We define the different OFF regions according to Table \ref{tab:obs_stra}. Given the stochastic nature of the emission of astrophysical gamma-ray photons and the detection of these by CTA, we create 50 simulated observations for the SHMF $\alpha=1.9$ DM model case and average the results in order to obtain statistically meaningful results. 

\begin{table}[h!]
\centering
\begin{tabular}{| c | c |}
\hline
N$_{obs}$ & 50 \\
\hline
T$_{obs}$ [h] & 300 \\
\hline
Obs. strategy & 1 ON / 3 OFF \\
\hline
Pointing $(l,\;b)$ [deg] & (150.57, -13.26) \\
\hline
Offset [deg] & 1.0\\
\hline
On Region [deg] & 1.0 \\
\hline
Energy range [TeV] & 0.03 - 100 \\
\hline
\end{tabular}
\caption{\label{tab:obs_stra} Observation parameters set-up for the Perseus DM-search campaign with the ON/OFF method.}
\end{table}

\begin{figure}[h!]
\centering
\includegraphics[width=0.49\linewidth]{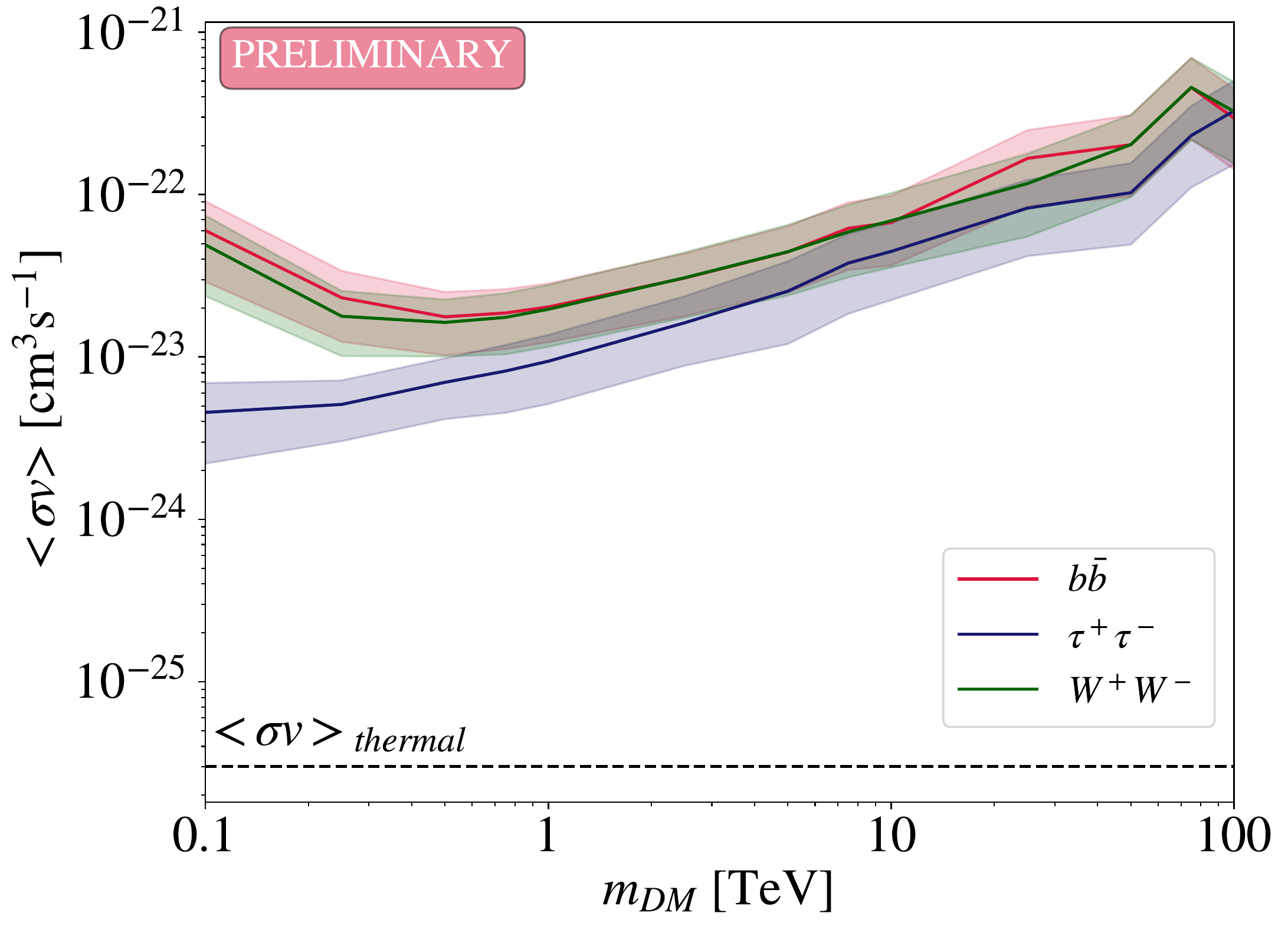}
\includegraphics[width=0.49\linewidth]{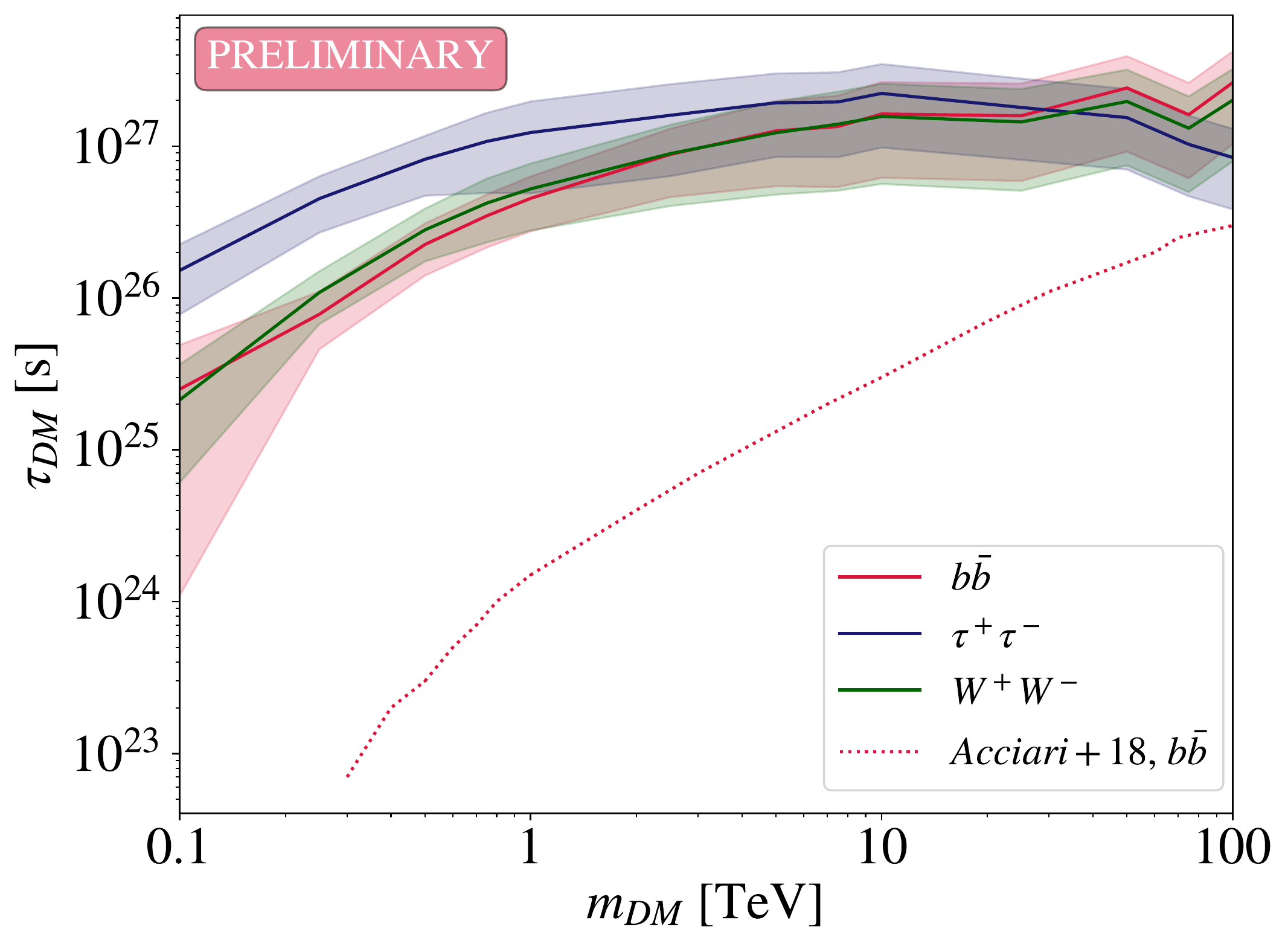} 
\caption{95\% C.L. upper limits on the DM parameter space. \textit{Left panel:} Averaged annihilation cross section versus DM mass. The thermal relic cross section is shown as a black dashed line. \textit{Right panel:} DM particle lifetime versus DM mass. For comparison, we show in red dotted line the constraints obtained from 202 h observation of Perseus for the $b\bar{b}$ channel with the MAGIC telescopes \cite{Acciari:2018sjn}.}
\label{fig:sigmav_vs_mass}
\end{figure}

As a first order approximation, we assume Perseus cluster to be a point-like source. No signal is found in these simulations, thus we compute the 95\% confidence level (C.L.) upper limits to the gamma-ray flux. Then, we proceed and compute 95\% C.L. upper limits to the DM annihilation cross section (Eq.~\ref{eqn:annihil-flux}) and decay lifetime (Eq.~\ref{eqn:decay-flux}). The preliminary results are shown in Figure~\ref{fig:sigmav_vs_mass} for the $b\bar{b}$, $\tau^+\tau^-$ and $W^+W^-$ channels. 

From the left panel of Figure \ref{fig:sigmav_vs_mass} we can see that our constraints to DM annihilation are, at best, around two orders of magnitude above the thermal relic cross section, even with the inclusion of halo substructure. Yet, for decay DM, the results of the right panel show that CTA will be able to improve the limits from \cite{Acciari:2018sjn} by up to two orders of magnitude in the lowest considered mass range. 

In the future, our analysis will necessarily consider the three known gamma-ray sources in Perseus, as well as the extension of the involved gamma-ray signals. An extended analysis, i.e. taking into account the extension of the cluster, is expected to worsen the obtained DM constraints. We will also include systematic uncertainties. An alternative analysis approach will also be performed via a template analysis fitting, as done for CTA in \cite{Acharyya:2020sbj}. The result of this analysis yield the first DM annihilation constraints for Perseus from any IACT, that are complementary to the ones obtained by the \textit{Fermi}-LAT Collaboration~\cite{Ackermann_2010, Zimmer:2015fup, Thorpe-Morgan:2020czg}.

\bibliographystyle{JHEP}
{\footnotesize\bibliography{biblio}}

\clearpage
\section*{Full Authors List: The Cherenkov Telescope Array Consortium July 2021 Authors}

H.~Abdalla\textsuperscript{1}, H.~Abe\textsuperscript{2},
S.~Abe\textsuperscript{2}, A.~Abusleme\textsuperscript{3},
F.~Acero\textsuperscript{4}, A.~Acharyya\textsuperscript{5}, V.~Acín
Portella\textsuperscript{6}, K.~Ackley\textsuperscript{7},
R.~Adam\textsuperscript{8}, C.~Adams\textsuperscript{9}
% ...

1 : Centre for Space Research, North-West University, Potchefstroom, 2520, South Africa

2 : Institute for Cosmic Ray Research, University of Tokyo, 5-1-5, Kashiwa-no-ha, Kashiwa, Chiba 277-8582, Japan

3 : Pontificia Universidad Católica de Chile, Av. Libertador Bernardo O'Higgins 340, Santiago, Chile

4 : AIM, CEA, CNRS, Université Paris-Saclay, Université Paris Diderot, Sorbonne Paris Cité, CEA Paris-Saclay, IRFU/DAp, Bat 709, Orme des Merisiers, 91191 Gif-sur-Yvette, France

5 : Centre for Advanced Instrumentation, Dept. of Physics, Durham University, South Road, Durham DH1 3LE, United Kingdom

6 : Port d'Informació Científica, Edifici D, Carrer de l'Albareda, 08193 Bellaterrra (Cerdanyola del Vallès), Spain

7 : School of Physics and Astronomy, Monash University, Melbourne, Victoria 3800, Australia

8 : Laboratoire Leprince-Ringuet, École Polytechnique (UMR 7638, CNRS/IN2P3, Institut Polytechnique de Paris), 91128 Palaiseau, France

9 : Department of Physics, Columbia University, 538 West 120th Street, New York, NY 10027, USA

% ...

%

%
%
%

\end{document}